\documentstyle[12pt]{article}
\catcode`\@=11
\def\makepreprititle{\par
  \begingroup
  \def\thefootnote{\fnsymbol{footnote}}
  \def\
@makefnmark{\hbox
  to 0pt{$^{\@thefnmark}$\hss}}
  \if@twocolumn
  \twocolumn[\@makepreprititle]
  \else \newpage
  \global\@topnum\z@
  \@makepreprititle \fi\thispagestyle{empty}\@thanks
  \endgroup
  \setcounter{footnote}{0}
  \let\makepreprititle\relax
  \let\@makepreprititle\relax
  \gdef\@thanks{}\gdef\@author{}\gdef\@title{}
  \gdef\@preprintnumber{}\gdef\@preprintdate{}\gdef\subtitle{}
  \let\thanks\relax}
\def\preprintnumber#1{\gdef\@preprintnumber{#1}}
\def\preprintdate#1{\gdef\@preprintdate{#1}}
\def\subtitle#1{\gdef\@subtitle{#1}}
\def\@makepreprititle{\newpage
{\def\baselinestretch{1}
  \begin{flushright} \@preprintnumber \par
  \@preprintdate \end{flushright} } \par
  \begin{center}
\vskip 1.5em
  {\LARGE \@title \par} \vskip 2.5em
  {\Large \lineskip .5em
  \begin{tabular}[t]{c}\@author
  \end{tabular}\par}
  \vskip 1em {\large \@date} \end{center}
  \par
  \vfil}
\date{\sl Theory Group, KEK, Tsukuba, Ibaraki 305, Japan}
\preprintdate{~}
\preprintnumber{~}
\subtitle{}
\def\abstract{\if@twocolumn
\section*{Abstract}
\else \normalsize
\begin{center}
{\bf Abstract\vspace{-.5em}\vspace{0pt}}
\end{center}
\quotation
\addtocounter{page}{-1}
\fi}
\def\endabstract{\if@twocolumn\else\endquotation\fi}
\def\spacing#1{\def\baselinestretch{#1}
\typeout{baselinestretch is modified to \baselinestretch}}
\catcode`\@=12

\spacing{1.4}
\preprintnumber{KEK-TH-365\\ KEK Preprint 93-68}
\preprintdate{July, 1993}
\title{Light Stop and the $b\rightarrow s\gamma$ Process}
\author{Yasuhiro~Okada}
\begin{document}
\makepreprititle
\begin{abstract}
A constraint from $b\rightarrow s\gamma$ process to the minimal
supersymmetric standard model is derived in the light stop region
where the coupling between the lighter stop and the $Z^0$ boson
vanishes due to the left-right mixing of the stop states.  It is
pointed out that although some region in the parameter space is
excluded from this process there remains a large parameter space where
the amplitude of the $b\rightarrow s\gamma$ is suppressed due to
partial cancellation between different diagrams.  Stop as light as 20
GeV is still viable from the $b\rightarrow s\gamma$ constraint.
\par
\vfil
\end{abstract}

Although supersymmetry (SUSY) has attracted much attention as
a promising candidate
of physics beyond the standard model of the elementary particle physics we
do not have any direct evidence of superpartners which are predicted to
exist in the supersymmetric theory. It is therefore very important to search
for any direct and indirect information on superpartners in order to test
the idea of supersymmetry.

In this respect the flavor changing neutral current (FCNC) process deserves a
special attention. In the supersymmetric standard model the FCNC is induced by
loop effects just as in the case of the standard model.
The origin of the flavor mixing is present not only
in the quark mass matrices but also in those of its superpartner, i.e. squark.
Therefore, FCNC process
is sensitive to the masses and mixings of the squark sector. In fact it is well
known that the squark masses are severely constrained from the $K^0-\bar{K}
^0$ mixing for general class of supersymmetric standard models
at least for the first two generations\cite{KK}. This is one of
motivations to consider so called minimal supergravity model where the
constraint from the $K^0-\bar{K}^0$ mixing can be avoided because all the
scalars have a common SUSY soft breaking mass at the GUT scale
\cite{donogh, KKSG}.

Another important FCNC process in the supersymmetric model
is the  $b\rightarrow s\gamma$ process.
It is well known that in the standard model the inclusive branching ratio
of this process is about $3$x$10^{-4}$ after taking account
of the QCD correction\cite{bsgsm,Misiak}.
This is compared with the recent
improved upper bound  $5.4$x$10^{-4}$ from the CLEO
collaboration\cite{CLEO}.  In the supersymmetric theory
we have four new contributions due to loops of (i) charged Higgs and up-type
quark, (ii) chargino
and up-type squark, (iii) gluino and down-type squark, (iv) neutralino and
down-type squark. In the context of the minimal supergravity model it is known
that the charged Higgs contribution as well as the chargino and gluino
contribution is sizable and, for some choice of parameters, can be
comparable to the standard model
contribution although the neutralino loop contribution (iv) is very small
\cite{bsgsusy}-\cite{Lopez}.

In this paper we study contribution to the $b\rightarrow s\gamma$
process from the light stop loop effect in the context of the minimal
supersymmetric standard model. Since the stop sector can have a
sizable left-right mixing it is possible that one of the two stop
states becomes much lighter than other squarks.  In fact it was
pointed out in Ref.~\cite{HK} that a stop lighter than 45 GeV is still
allowed since we can arrange the squark mixing so that the lighter
stop does not couple to the $Z^0$ boson.  In such case the
experimental lower bound of the stop mass is given from the TRISTAN
experiments. The bound is about 27 GeV provided that the photino mass
is not close to the stop mass \cite{tristan}.  It is also possible
that such a light stop can escape detection at the Fermilab Tevatron
as discussed in Ref.~\cite{Baer}.  If we assume that the stop exists
in such a low mass region we can expect that the $b\rightarrow
s\gamma$ amplitude becomes in principle very large because the stop
and chargino loop contribution can dominate this process, therefore
the improved measurement can put a useful constraint to the relevant
parameter space.  This is especially important if we assume the GUT
relations among the gaugino masses, since not only one of the stops
but also at least one of the charginos is necessarily light.  We
investigate the contribution from (ii) in the light stop region taking
account of the experimental constraints in the gaugino and higgsino
parameter space. We will see that although some parameter space is
excluded from the $b\rightarrow s\gamma$ measurement there still
remain an region where a stop as light as 20 GeV is not constrained.
This is due to a partial cancellation between two graphs involving the
top Yukawa and the SU(2) gauge coupling constants.

We consider here the usual minimal supersymmetric standard model with three
families of quarks and leptons and their supersymmetric partners. For the
parameters of the gaugino and higgsino sector we assume the GUT condition for
the Majorana mass terms of the gauginos, then three gaugino mass parameters
are related to each other in terms of gauge coupling constants. For FCNC
processes the squark mass matrices are especially important. We take
here that the
following simplified form for the up-type squark mass matrix \cite{oshimo}:

\begin{eqnarray}
\nonumber
   L_{u-squark} &=& (\tilde{u}_L~~\tilde{u}_R)^*
        \left( \begin{array}{cc}
		V_L(u) & 0\\
		0 & V_R(u)\end{array} \right)\\
\nonumber
     ~ &~&      \left( \begin{array}{cccccc}
		m_{u_L}^2 & ~ & ~ & ~ & ~ & ~  \\
		~ & m_{c_L}^2 & ~ & ~ & ~ & ~  \\
		~ & ~ & m_{t_L}^2 & ~ & ~ & am_t \\
		~ & ~ & ~ & m_{u_R}^2 & ~ & ~  \\
		~ & ~ & ~ & ~ & m_{c_R}^2 & ~  \\
		~ & ~ & a^*m_t & ~ & ~ & m_{t_R}^2
                              \end{array} \right) \\
     ~ &~&   \left( \begin{array}{cc}
		V_L^{\dag}(u) & 0\\
		0 & V_R^{\dag}(u)\end{array} \right)
	     \left( \begin{array}{c}
	           \tilde{u}_L \\ \tilde{u}_R
	          \end{array} \right),
\label{upmix}
\end{eqnarray}
where $V_L(u)$ and $V_R(u)$ are up-type quark mixing matrices, i.e.,
\begin{equation}
V_R^{\dag}(u) \cdot m_u \cdot V_L(u) = diagonal.
\end{equation}
In this form we have assumed that, apart from the left-right mixing
term for the third generation, the up-squark mass matrix is diagonal
in the  basis where the corresponding quark is diagonal. If we solve
the renormalization group equations of the squark mass matrices with
a universal scalar mass at the GUT scale the above form is obtained
as an approximate solution when one neglects the small off-diagonal
components for the first and the second generations
\cite{KKSG}. In the super GIM basis where $\tilde{u}'_L=V^{\dag}
_L(u)\tilde{u}_L,\tilde{u}'_R=V^{\dag}_R(u)\tilde{u}_R$
only the stop sector has a mixing,
\begin{equation}
L_{stop}=(\tilde{t}'_{L}~~\tilde{t}'_{R})
	\left( \begin{array}{cc}
	    m_{t_L}^2 & a m_t \\
	    a^* m_t & m_{t_R}^2 \end{array} \right)
	\left( \begin{array}{c}
	    \tilde{t}'_{L}\\
	    \tilde{t}'_{R} \end{array} \right),
\label{stopmix}
\end{equation}
then the light and heavy stop eigenstates are given by
\begin{eqnarray}
\nonumber
\tilde{t}_1 &=& \cos\theta_t\tilde{t}'_L-\sin\theta_t\tilde{t}'_R,\\
\tilde{t}_2 &=& \sin\theta_t\tilde{t}'_L+\cos\theta_t\tilde{t}'_R.
\end{eqnarray}
Here we have assumed that the coefficient $a$ is real for simplicity.
Since the coupling of the light stop $\tilde{t}_1$ to the $Z^0$ boson
is proportional to
\begin{equation}
\frac{1}{2}\cos ^2\theta_t-\frac{2}{3}\sin^2\theta_W,
\end{equation}
this coupling vanishes when $|\cos \theta_t| \sim 0.55$. In this region
even the stop lighter than 45 GeV is still allowed. Hereafter we will
concentrate our consideration to this case. Notice that there are two
distinct possibilities in which the above coupling vanishes, i.e.
$\tan \theta_t$ is positive or negative. This ambiguity comes from
the two choices of the sign in the off-diagonal term in Eq.(\ref{stopmix}).

The calculation of $b\rightarrow s\gamma$ branching ratio in the standard
model is given in Ref.~\cite{bsgsm,Misiak} and the SUSY contributions to
this process are described in Ref.~\cite{bsgsusy}. In the standard model
we first determine the weak effective Hamiltonian at the weak scale.
\begin{equation}
H_{eff}=-\frac{4G_F}{\sqrt{2}}V^*_{ts}V_{tb}\sum_{i=1}^{8}
       C_i(\mu)O_i,
\end{equation}
where the operators $O_i$'s are shown in Ref.~\cite{bsgsm} and $V_{ij}$
is the $i,j$ component of the Kobayashi-Maskawa matrix. Especially, the
$O_7$ is a magnetic moment operator given by
\begin{equation}
O_7=\frac{e}{16\pi^2}m_b\bar{s}_L\sigma^{\mu \nu}b_R F_{\mu \nu},
\end{equation}
and $O_8$ is a gluonic operator,
\begin{equation}
O_8=\frac{g}{16\pi^2}m_b\bar{s}_L\sigma^{\mu \nu}b_R G_{\mu \nu},
\end{equation}
where g is a strong coupling constant. The coefficients $C_7(m_W)$
and $C_8(m_W)$ are given from one loop calculation of
$b\rightarrow s\gamma$ diagrams.
The $b\rightarrow s\gamma$
amplitude is determined from the effective Hamiltonian at~$\mu=m_b$
scale, and the QCD correction is taken into account by solving the
renormalization group equations for the coefficient functions  $C_i(\mu)$
from the $m_W$ scale to the $m_b$ scale.
The inclusive branching ratio of $b\rightarrow s\gamma$ is then given by
\begin{equation}
Br(b\rightarrow s\gamma)=\frac{\Gamma(b\rightarrow s\gamma)}
      	{\Gamma(b\rightarrow ce\bar{\nu})}
			     Br(b\rightarrow ce\bar{\nu}),
\end{equation}
where $\Gamma(b\rightarrow s\gamma)$ is determined from the effective
Hamiltonian:
\begin{equation}
\Gamma(b\rightarrow s\gamma)=\frac{m_b^5}{64\pi^2}\frac{\alpha_W^2\alpha}
       {m_W^2}|V^*_{ts}V_{tb}|^2|C_{7eff}(m_b)|^2,
\end{equation}
where $C_{7eff}(m_b)=C_7(m_b)-\frac{1}{3}(C_5(m_b)+3C_6(m_b))$
\cite{Misiak}
and $\Gamma(b\rightarrow ce\bar{\nu})$ is the semi-leptonic
decay width of the b quark. The SUSY and other short distance
effects give extra contributions to the  $C_7(m_W)$.
\footnote{Strictly speaking, the initial condition of the coefficients
is given at the scale of the particle's mass inside the loop, not at
the $m_W$ scale. However, the correction is small
since we are interested in the case in which these masses are not very
different from $m_W$.}
The contribution from the chargino and stop
loop depends on the masses and mixings of the up-type squark sector
given by Eq. (\ref{upmix}) and the parameters of the chargino mass matrix, i.e.
the SU(2) gaugino mass parameter $M_2$, the higgsino mass parameter
$\mu$, and the vacuum angle $\tan\beta$. In the actual calculation, we have
taken into account the 8x8 anomalous dimension matrix of Ref.~\cite{Misiak}
and the QCD $\Lambda$ parameter $\Lambda_{LLA}^{f=5}=100$MeV and $m_b=4.5$GeV.
We calculated the
amplitude in the parameter space of $M_2$ and $\mu$ for different values
of the stop mass.

In Fig.1 we show the ratio of the $b\rightarrow s\gamma$ amplitude
of the chargino contribution to that of the standard model contribution.
For the standard model contribution we take the top quark mass to be 150 GeV.
The light
stop mass is taken to be 40 GeV and the stop mixing angle is fixed in such
a way that the stop and $Z^0$ coupling vanishes and $\tan\beta=2$. The
Fig.1 (a) and (b)
correspond to two choices of $\tan\theta_t$. The heavy stop mass and other
up-type squark masses are taken to be 200 GeV.
Since the amplitude is dominated by
the lightest stop loop the result does not depend strongly on the choice
of the other squarks' masses. In the figures we have shown the excluded
region in the $\mu-M_2$ space from the chargino and neutralino search
experiments at LEP\cite{LEPpr}.  The conditions which we take here to determine
the excluded region (I) are that the chargino mass is less than 45 GeV,
the invisible width of the $Z^0$ boson to the lightest
neutralino($\chi$) pair is more than 22MeV, and
the branching ratio of $Z^0 \rightarrow \chi \chi'$ and
$Z^0 \rightarrow \chi' \chi'$ is larger than $5$x$10^{-5}$
where $\chi'$ represents the neutralino other than the lightest one.
Also we have excluded the region (II) where the lightest stop is lighter than
the lightest neutralino which is assumed to be
the lightest supersymmetric particle (LSP). We see that although the
amplitude becomes large in some parameter region there remains a large
parameter space where the ratio is within $\pm$25\% of the standard model
amplitude especially for the choice of $\tan\theta_t>0$. In fact, in order
to just satisfy the present CLEO's bound ($5.4$x$10^{-4}$)  the amplitude from
the chargino graph can be $-2.4\sim0.4$ times the standard model amplitude,
therefore most of the parameter
space in the figures is allowed. The situation would change if the
experiment gives a finite value to the branching ratio since we can now
exclude a large parameter space where the standard model and chargino
contribution cancel each
other. Even in that case, however, no strong constraint is obtained
if the $\mu$ and $M_2$
are in the region where the amplitude is the same as that of the standard
model within $\pm25\%$ .  This is true even if we take the stop
mass as small as 20 GeV as shown in Fig. 2. Here we take  $\tan \beta =1.8$
and $\tan \theta_t >0$ .
In this case the phenomenological
allowed region is confined to the negative $\mu$ region where
the chargino contribution to the
$b\rightarrow s\gamma$ amplitude is also suppressed and the ratio is about
-0.25 in the most of the parameter space shown here. For $\tan \theta_t <0$
this ratio is $+0.25\sim +0.5$ in the same region which corresponds to
a branching ratio  $1.5\sim 2.2$ times as large as  that of the standard model.
\footnote{If the stop is as light as 20 GeV, the TRISTAN experiment becomes
important. However, the sensitivity of the stop search in the single photon
annihilation process is lost if the mass
difference between the stop and the LSP neutralino is within a few GeV\cite
{tristan}.
The author thanks R. Enomoto for explaining this point. }

We can understand this property by looking at the light stop contribution
to $C_7(m_W)$,
\begin{eqnarray}
\nonumber
C_7 &=& \frac{m_W^2}{m_{\tilde{t}_1}^2}\sum_{j}\{| V_{j1}\cos\theta_t+V_{j2}
z_t\sin\theta_t|^2 f^{(1)}(x_{\chi_j\tilde{t}_1})\\
  ~&~& -( V_{j1}\cos\theta_t+V_{j2}z_t\sin\theta_t)U_{j2}\cos\theta_t
	z_{\chi_j} f^{(2)}(x_{\chi_j\tilde{t}_1}) \},
\label{C7}
\end{eqnarray}
where
\begin{eqnarray}
f^{(1)}(x_{\chi_j \tilde{t}_1}) &=& F_1(x_{\chi_j \tilde{t}_1})+\frac{2}{3}
	     F_2(x_{\chi_j \tilde{t}_1}),\\
f^{(2)}(x_{\chi_j \tilde{t}_1}) &=& F_3(x_{\chi_j \tilde{t}_1})+\frac{2}{3}
	     F_4(x_{\chi_j \tilde{t}_1}),
\end{eqnarray}
\begin{equation}
z_t=\frac{m_t}{\sqrt{2}m_W\sin\beta},~~z_{\chi_j}=\frac{m_{\chi_j}}
      {\sqrt{2}m_W\cos\beta},
\end{equation}
\begin{eqnarray}
F_1(x) &=& \frac{1}{12(x-1)^4}(x^3-6x^2+3x+2+6x\log x),\\
F_2(x) &=& \frac{1}{12(x-1)^4}(2x^3+3x^2-6x+1-6x^2\log x),\\
F_3(x) &=& \frac{1}{2(x-1)^3}(x^2-4x+3+2\log x),\\
F_4(x) &=& \frac{1}{2(x-1)^3}(x^2-1-2x\log x),
\end{eqnarray}
and $U$ and $V$  are matrices which diagonalize the chargino sector.
\begin{eqnarray}
U^*M_CV^{\dag}&=&\left( \begin{array}{cc}
                m_{\chi_1} & 0 \\
                0 &  m_{\chi_2}\\
                \end{array} \right) ,\\
M_C &=&  \left( \begin{array}{cc}
                M_2 & \sqrt{2}m_W \sin \beta\\
                \sqrt{2}m_W \cos \beta & \mu\\
                \end{array} \right) .
\end{eqnarray}
We have used an abbreviation $ x_{\chi_j \tilde{t}_1}= \frac{m_{\chi_j^2}}
{m_{\tilde{t}_1}^2}$.
Numerically,$f^{(2)}$ is much larger than $f^{(1)}$, therefore the second
term in Eq. (\ref{C7}) is more important.
The two terms in the second term, i.e.
$V_{j1}\cos\theta_t$ and $V_{j2}z_t\sin \theta_t$ correspond to two diagrams
which depend on the SU(2) gauge coupling constant and the top Yukawa
coupling constant respectively as shown in Fig. 3.
When $\tan \theta_t>0$ the two contributions tend to cancel each other,
then the stop contribution is suppressed. In fact a complete cancellation
occurs along a line which passes near the origin of the $\mu$- $M_2$ space.
On the other hand
when $\tan \theta_t<0$ we don't see such cancellation and the amplitude
is large in most of the parameter space.

We have calculated the $b\rightarrow s\gamma$ amplitude in the
light stop region where the coupling between the $Z^0$ and the light stop
is suppressed. We see that the contribution to the $b\rightarrow s\gamma$
process is also suppressed in a large parameter region for one choice of
$\tan \theta_t$ and therefore the light stop as light as 20 GeV is still
allowed from the  $b\rightarrow s\gamma$ constraint. It is known
that in the SUSY model other contributions such as the charged Higgs or
gluino loop gives sizable effects to the amplitude. Since the chargino
and gluino contributions can have either sign for the amplitude relative
to the standard model and the charged Higgs contribution, it is possible
that different SUSY contributions cancel each other and in such cases
relatively light masses for the charged Higgs and/or SUSY particles are
allowed. What is remarkable in the light stop case is, however, that the
chargino contribution itself is suppressed without relying on any cancellation
with other SUSY contributions. \\

The author would like to thank R. Enomoto, K. Hagiwara, K. Hikasa,
M. Kobayashi, A.I. Sanda, and T. Yanagida for useful discussions
and comments. This work is supported in part by the Grant-in-aid for
Scientific Research from the Ministry of Education, Science and Culture
of Japan.

\newpage
\noindent{\large {\bf Figure Captions}}\\
\\
Fig.1 ~~The chargino loop contribution to the $b\rightarrow s\gamma$
amplitude normalized to the standard model amplitude in the $M_2-\mu$
space. The numbers in x and y axes are in GeV. The light stop mass
is taken to be 40 GeV and other squarks' masses are 200 GeV and $\tan\beta=2$.
(a) and (b) correspond to the stop mixing angle $\tan\theta_t=1.51$ and
$\tan\theta_t=-1.51$ respectively. I is the region excluded from
the chargino and neutralino search at LEP and II is the region where
the stop becomes lighter than the lightest neutralino.\\
{}~\\
Fig.2 ~~The same contour plot as Fig. 1 in the case that the light stop
is 20 GeV and  $\tan \beta$=1.8 and $\tan\theta_t=1.51$.
Other parameters and the meanings
of the excluded region I, II are the same as in Fig. 1.\\
{}~\\
Fig.3 ~~Two diagrams which contribute to the $b\rightarrow s\gamma$ amplitude.
$y_b,~y_t,$ and $g_2$ represent the bottom Yukawa, top Yukawa and SU(2)
gauge coupling constants respectively. The photon $(\gamma)$ line can
be attached either to the stop $(\tilde{t}_1)$ or the chargino $(\chi^-)$
line.

\end{document}